\documentclass[reprint,amsmath,amssymb,
 aps,prb,showpacs]{revtex4-1}
\usepackage{graphicx}
\usepackage{dcolumn}
\usepackage{bm}
\begin{document}

\title{Effect of Epitaxial Strain on Phase Separation in Thin Films}

\author{Arka Lahiri}
 \author{T.A. Abinandanan}%
\affiliation{%
Department of Materials Engineering, 
Indian Institute of Science, Bangalore - 560 012, India.
}%

\author{M.P. Gururajan}
 \affiliation{
Department of Metallurgical Engineering and Materials Science, Indian Institute of Technology - Bombay, Mumbai - 400076, India.
}%
\author{Saswata Bhattacharyya}
\affiliation{%
Department of Material Science and Engineering, 
Indian Institute of Technology - Hyderabad, Hyderabad - 502205, India.
}%
\date{\today}

\begin{abstract}
We examine the role of an imposed epitaxial strain $e$ in enhancing or depressing the spinodal instability of an alloy thin film. Since the alloy film starts with an imposed strain, phase separation offers a mechanism to relieve it, but only when the film is elastically inhomogeneous. With composition-dependence of elastic modulus given by $y$, and that of lattice parameter by $\eta$, our simulations using the Cahn-Hilliard model show (and analytical results for early stages confirm) that, for $(ey /\eta) >0$, the imposed strain adds to the driving force for phase separation, decreases the maximally growing wave length, and expands the coherent spinodal in the phase diagram. Further, when $(ey/\eta) > 0.372$, it expands to even outside of chemical spinodal. Phase separation produces islands of elastically softer (harder) phase with (without) a favorable imposed strain. These results are in agreement with experimental results in GeSi thin films on Si and Ge substrates, as well as in InGaAs films on GaAs  substrates.
\end{abstract}


\maketitle
Phase separation via spinodal decomposition in an epitaxial thin film is exploited to produce self-assembled quantum dots or wires in systems such as InGaAs/GaAs (i.e., InGaAs films on a GaAs substrate)~\cite{Liu2000,Adhikary2012},
GeSi/Si~\cite{Brunner2002,Aqua2013},
Ge-Sn/Ge~\cite{Ragan2006,Ragan2005}, InAlAs/InP~\cite{Twesten1999}, 
and InGaP/GaAs~\cite{Bortoleto2007}. The epitaxial strain imposed on the film has a strong effect on enhancing or depressing this spinodal instability: while GeSi/Si films under a compressive  strain produce Ge-rich quantum dots~\cite{Brunner2002,Aqua2013}, GeSi/Ge films under a tensile strain resist phase separation ~\cite{Myronov2011,LiuX2011}. This paper addresses this asymmetric effect of the epitaxial strain on phase separation, both theoretically and through simulations. 

Following the classic work of Asaro and Tiller~\cite{Asaro_tiller1972} and Grinfeld~\cite{Grinfeld1986} on stress induced surface modulations, morphological instability in stressed thin films has been studied in single component films~\cite{Srolovitz1989,Spencer1991} and alloy films~\cite{Leonard1998,Huang2002, Guyer1995,Spencer2001,Desai2010}.  However, as we show in this communication, spinodal decomposition, by itself, can provide a way to relieve the strain energy in epitaxial alloy thin films. 

Elastic stress effects on spinodal decomposition have been studied extensively (see Ref.~\cite{Voorhees2004} for reviews), starting with Cahn's classic work~\cite{Cahn1961} which examined the spinodal instability of an alloy with a lattice parameter mismatch, $\eta$, but with  a composition-independent modulus, and showed that strains due to a finite $\eta$ enhance the alloy's stability; the coherent spinodal is submerged within the chemical spinodal. Later studies considered systems with a composition-dependent modulus under no stress ~\cite{Onuki1991}, a hydrostatic stress and under a uniaxial stress~\cite{Thompson1997} in the limit of a small modulus mismatch. 

Epitaxial thin films are an excellent model system for exploring elastic stress effects, because epitaxial strains as large as 5 \% may be imposed on them without triggering plastic deformation. Since these films start with considerable elastic strain, phase separation may relieve (or add to) the strain energy depending on the sign of elastic inhomogeneity relative to the sign of the imposed strain to which it couples. Such a coupling, therefore, may promote or suppress spinodal instability.
 
We model phase separation  in an alloy thin film under an imposed (epitaxial) strain using the Cahn-Hilliard equation~\cite{Cahn1961}:  
\begin{eqnarray}
\frac{\partial c}{\partial t} = \nabla \cdot M\nabla {\mu},
\label{Cahn_Hilliard}
\end{eqnarray}
where $M$ is the mobility and $\mu$ is the chemical potental. 
The film is in mechanical equilibrium at all times:
\begin{eqnarray}
\nabla \cdot \sigma^{el}=0
\label{mech_eq}
\end{eqnarray} 
where $\sigma^{el}_{ij} = C_{ijkl}\epsilon^{el}_{kl}$ is the elastic stress tensor, $\epsilon^{el}_{kl}$ is the elastic strain field, and $C_{ijkl}$ is the composition-dependent elastic modulus.  

Chemical potential $\mu=\delta(F/N_{V})/\delta c$ is obtained from a free energy functional which includes a sum of chemical and elastic energy densities: 
\begin{eqnarray}
F=\int_\Omega \left[ f_{0}(c)+\kappa {(\nabla c)}^{2}  + \frac{1}{2}\sigma_{ij}^{el} \epsilon_{ij}^{el} \right]d\Omega \, ,
\label{free_energy}
\end{eqnarray}
where $\kappa$ is the gradient energy coefficient, $f_{0} = N_{V} \, A\, c^{2}(1-c)^{2}$ is the bulk free energy density in its usual double well form, with $A$ setting the barrier between the two equilibrium phases, and $N_{V}$ is the number of molecules per unit volume. 

The elastic strain is given by $\epsilon_{ij}^{el}=\epsilon_{ij}-\epsilon_{ij}^{0}$, where   $\epsilon_{ij}^{0} = (c-c_{0})\eta\delta_{ij}$ is the composition-dependent eigenstrain, $\epsilon_{ij} = (u_{i,j} + u_{j,i}) / 2$ is the strain field obtained from the deformation field $u_{i}$, and $\delta_{ij}$ is the Kronecker's delta (identity) tensor. $\eta=\left[(1/a)(da/dc) \right]_{c=c_{0}}$ is the Vegard's law coefficient: $a$ is the composition-dependent lattice parameter, $a_{0} = a(c_{0})$ and $c_{0}$ is composition of the undecomposed alloy. In the technique employed for solving Eq.~\ref{mech_eq}, we have $\epsilon_{ij}=e \delta_{ij}+ \epsilon_{ij}^{*}$, where $e\delta_{ij}$ and $\epsilon_{ij}^{*}$ denote the epitaxially imposed homogeneous strain field and the heterogeneous strain field~\cite{Khachaturian1983} respectively. Assuming elastic isotropy, and Vegard's law behavior for the Young's modulus $Y(c)$, we have: 
\begin{eqnarray}
Y(c) = Y_{0}\left[1 + (c-c_{0}) y \right] ,
\label{inhom_mod_par}
\end{eqnarray}
where $Y_{0} = Y(c_{0})$, and $y=\left[(1/Y)(dY / dc)\right]_{c=c_{0}}$ is the Vegard's law coefficient for the modulus. 

Thus, the elasticity-related parameters in our model are the Vegard's law coefficients $\eta$ and $y$, and the epitaxial strain parameter $e$. 

Following Cahn, we solve a linearized version of Eq.~\ref{Cahn_Hilliard}  (valid for early stages when $\partial^{2}f_{0} / \partial c^{2}$ may be assumed to be a constant), and derive an expression for $\beta_{m}$, the wave number of the composition modulation with the fastest growing amplitude. For solving Eq.~\ref{mech_eq} (to compute the contribution of strain energy towards $\mu$), we took inspiration from the iterative solution technique used in the simulations; thus, we perform the zeroth and first iteration of the technique analytically to compute the strain field  due to a composition wave. This first-order solution to the linearized Cahn-Hilliard equation yields the following expression for the maximally-growing wave number (details will be published elsewhere): 
\begin{widetext}
\begin{eqnarray}
\beta_{m}^{2}\; = \; \frac{- \left( \partial ^{2} f_{0} / \partial c^{2} \right) - Y_{0} \eta^{2} + 2 Y_{0}ey\eta + (Y_{0}{(1+\nu)}^{2} y^{2}e^{2}/(1-\nu^{2}))}{4\kappa N_{V}} \nonumber \\
\label{betamax}
\end{eqnarray}
\end{widetext}
where $Y_{0}/2$ is the 2D plane stress analog of $Y_{0}/(1-\nu)$ derived by Cahn for 3D.

We have also solved Eq.~\ref{Cahn_Hilliard} numerically to simulate phase separation in a thin film under periodic boundary conditions, using an iterative Fourier spectral technique~\cite{Khachaturyan1995,Michel1999,Hu2001,Gururajan2007})
for solving Eq.~\ref{mech_eq}, a semi-implicit Fourier spectral technique~\cite{Chen1998} for solving Eq.~\ref{Cahn_Hilliard}, and the FFTW package~\cite{Frigo2005} for  computing the discrete Fourier transforms. From these simulations, we have extracted $\beta_{m}$ for early stages of phase separation.

\begin{figure}[h]
\includegraphics[scale=0.7]{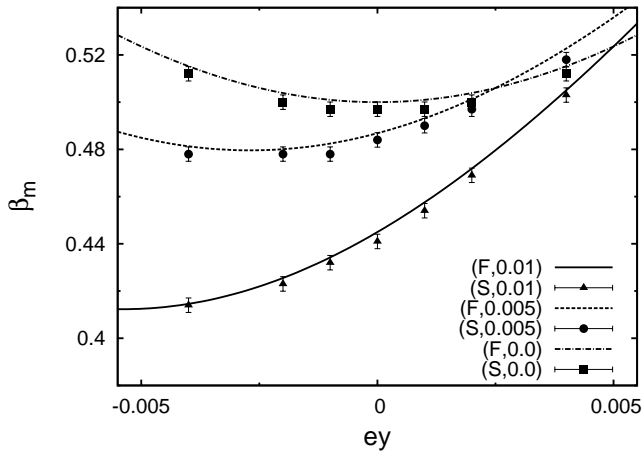}
\caption{$\beta_{m}$ vs $ey$ for $\eta=0.0,0.005,0.01$; In the figure legend `S' stands for results obtained from simulations, while `F' denotes the analytical solution: the numerical value following `S' or `F' denotes the value of $\eta$. The simulation parameters used are:$A=1$, $M=1$, $\kappa=1$,  $Y_{0}/N_{V}=2080$, $\nu=0.3$. We used a $2048\times2048$ simulation cell with $dx=dy=1$. A $dt$ of $0.1$ was chosen as our timestep size and a total time till $10$ was considered to be early stage. The homogeneous alloy composition was $c_{0}=0.5$.The simulation parameters were non-dimensionalised \cite{Gururajan2007}. Due to discrete nature of the Fourier space\cite{Press1992}, the error associated with the measurement of $\beta_{m}$ from our simulations is $0.00306$.}
\label{F1}
\end{figure}
 
In Fig.~\ref{F1}, we compare the wave number $\beta_{m}$ of the fastest-growing composition wave from simulations with that obtained analytically. The agreement between the two is excellent, indicating that the first order solution to Eq.~\ref{mech_eq} is sufficient to account for the simulation results from early stages. Since Fig.~\ref{F1} plots $\beta_{m}$ against $ey$, the data points obtained for $ey = 0$ for various values of $\eta$ correspond to the solution obtained by Cahn~\cite{Cahn1961}. 

The results in Fig.~\ref{F1} may be understood in terms of two factors: first, when $\eta=0$ (i.e., the second and third terms in Eq.~\ref{betamax} are zero, and the fourth term is proportional to $(ey)^{2}$), $\beta_{m}$ rises symmetrically with $|ey|$. In other words, a finite $e$ adds to the driving force for phase separation. This is explained by elastic strain (which is also the same as the total strain) being smaller at stiffer regions, and larger at softer regions; this combination relieves some of the stored strain energy, and leads to an increase in $\beta_{m}$. 
 
The second factor is that systems with a finite eigenstrain (i.e., $\eta > 0$) have non-zero values for the second and third terms in Eq.~\ref{betamax}. Specifically, the third term is linear in $ey$, and therefore, leads to an asymmetry in the curves for finite $\eta$ in Fig.~\ref{F1}. When $(ey/\eta) > 0$,  it favors phase separation and increases $\beta_{m}$ by accentuating the difference between the strains at stiffer and softer regions: with a finite $\eta$, stiffer (softer) regions are at an even smaller (larger) strain than with $\eta=0$. Thus, the curves in Fig.~\ref{F1} for $\eta > 0$ show a positive slope at $ey = 0$. 
	
Since Eq.~\ref{betamax} is quadratic in $ey$, $\beta_{m}$ has a minimum at $(ey/\eta) = - 0.538$, in good agreement with  simulation results.  

\begin{figure}[h]
\includegraphics[scale=0.7]{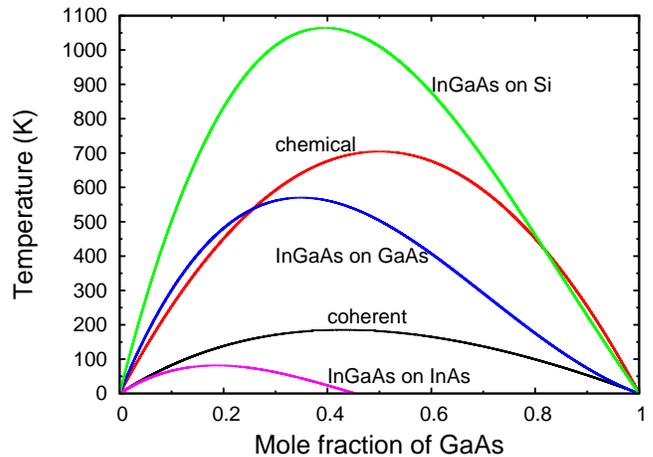}
\caption{(Color online)The chemical, coherent and the constrained spinodal lines for InGaAs on GaAs, InAs and Si. The parameters which were used were: $\Omega=11715.2$\,J/mol  \cite{Stringfellow1969}, and $Y_{InAs}=103$\,GPa, $Y_{GaAs}=154.4$\,GPa\cite{Adachi2009}, $\nu=0.3$, $V_{InAs}=33.4 \times 10^{-6}
{\text{m}}^{3}\text{/mol}$ and $V_{GaAs}=27.2 \times 10^{-6}          {\text{m}}^{3}\text{/mol}$.The lattice parameters used were: $a_{InAs}=0.6058$\,nm, $a_{GaAs}=0.5653$\,nm and $a_{Si}=0.543$\,nm \cite{Adachi2009}}
\label{F3}
\end{figure}

Another feature of our work is that when $(ey/\eta) > 0.372$ (and also when $(ey/\eta)<-1.449$), the sum of all the elasticity-related terms (i.e. the second, third and fourth terms) in Eq.~\ref{betamax} is positive. Physically, this condition implies that strain relief due to phase separation adds to the driving force for spinodal decomposition; the coherent spinodal curves for $(ey/\eta) > 0.372$ may extend to regions even outside the chemical spinodal (see discussion on Fig.~\ref{F3}, below).

\begin{figure}
\includegraphics[scale=0.6]{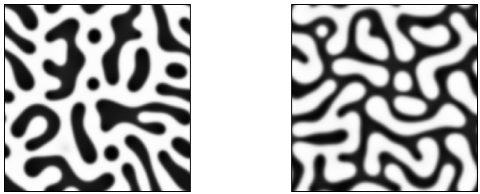}
\caption{Microstructures at a total time of 1400, under epitaxial strains of $e=0.02$ and $e=0.0$ for the images on the left and right respectively. The other parameters used in the simulation are:$A=1$, $M=1$, $\kappa=1$,  $Y_{0}/N_{V}=2080$, $\eta=0.01$, $\nu=0.3$, $y=0.4$, $c_{0}=0.5$.}
\label{F2}
\end{figure}

Of course, simulations based on the Cahn-Hilliard model are not restricted to the early stages. And, once again, the epitaxial strain has an interesting effect on the late stage microstructures in Fig.~\ref{F2}. When $ey/\eta > 0$, phase separation is enhanced, and the left figure shows elastically softer phase (the darker phase) as islands embedded in a matrix of the elastically harder (lighter) phase. The microstructure on the right, on the other hand, is for $ey = 0$; it shows islands of the harder phase embedded inside a matrix of the softer phase, consistent with the results of Onuki~\cite{Onuki1991}.

By generalizing these results to other temperatures and alloy compositions, we can compute the limit of spinodal instability in a phase diagram. For this purpose, we have chosen the GaAs-InAs pesudobinary system; with a regular solution model for $f_{0}$, and neglecting the temperature dependence of the elastic moduli and lattice parameters of InAs and GaAs, a linear stability analysis of Cahn~\cite{Cahn1961} yields the following expression for the coherent spinodal curves for a thin film under an epitaxial strain $e$:
\begin{eqnarray}
\frac{\partial ^{2} f_{0}}{\partial c^{2}} + \left[\eta^{2}Y_{0} \left(1-\frac{2ey}{\eta} \right) -\frac{Y_{0}{(1+\nu)}^{2} y^{2}e^{2}}{(1-\nu^{2})}\right]V=0\nonumber \\
\label{spinodal}
\end{eqnarray}
where $V$ denotes the molar volume. [We note that we recover Cahn's expression for the coherent spinodal (Eq.14 in Ref.~\cite{Cahn1961}) by setting $y=0$ and replacing $Y_{0}$  by $2Y_{0}/(1-\nu)$ appropriate for 3D systems].

In Fig.~\ref{F3}, we show the chemical spinodal curve obtained by setting $(\partial^{2}f_{0} / \partial c^{2}) = 0$; for the regular solution model used here, this curve is symmetric about $c=0.5$. Since the elastic modulus depends on composition, the coherent spinodal is asymmetric -- the curve is closer to the chemical spinodal on the softer InAs-rich side (i.e., $y > 0$).

In Fig.~\ref{F3} we also show the coherent spinodals for InGaAs films grown on GaAs, InAs as well as Si. For those on InAs, a tensile epitaxial strain (i.e., $(ey / \eta) < 0$) suppresses spinodal demposition, and the coherent spinodal for InGaAs/InAs is  well below that for no epitaxial strain. In contrast, spinodal decomposition is enhanced in InGaAs films on GaAs since ($ey/\eta) > 0$. This asymmetry in the coherent spinodal is in qualitative agreement with that in the miscibility gap reported in Ref.~\cite{Johnson1988} for a similar system.

Finally, for films grown hypothetically on Si (hypothetically, because InAs-rich films on Si would suffer unreasonably large epitaxial strains of 10 \% or more), conditions for spinodal decomposition are so much more favorable than for those on GaAs  that the spinodal curve expands to regions even outside the chemical spinodal. 

Consistent with these results, there is ample evidence for phase separation in InGaAs films grown on GaAs (see  Ref.~\cite{Skolnick2004,Bhattacharya2004} for reviews). Our analysis also predicts a suppression of spinodal decomposition for these films when grown on an InAs substrate.

We now turn to SiGe films. Even though the critical temperature for the chemical spinodal is around 300 K (and that of coherent spinodal is even lower), Ge-rich quantum dots are formed in SiGe/Si films Si\cite{Brunner2002,Aqua2013} at growth temperatures of over 700 K, indicating surface thermodynamic origins of the driving force for phase separation. Treating Ge as the solute, we have $\eta > 0$, $y < 0$; therefore, phase separation is promoted on a Si subtrate (since $(ey / \eta) > 0$). More importantly, phase separation is suppressed on a Ge substrate (for which $(ey / \eta) < 0$) This is indeed seen in SiGe films grown on Ge \cite{Myronov2011,LiuX2011}.

Our work is similar in spirit to that of Onuki and Thompson and Voorhees who studied the effect of elastic inhomogeneity on spinodal decomposition for systems with a small mismatch in modulus. While the former studied systems under no stress~\cite{Onuki1991}, Thompson and Voorhees~\cite{Thompson1997} studied systems under a hydrostatic and uniaxial stress. Our analytical results for $\beta_{m}$, on the other hand, involve no approximation about the modulus mismatch (i.e., $y$ can have any value). 

Finally, we have considered phase separation within the plane of a thin film. However, when the process occurs via surface diffusion during growth, islands of the soft phase can grow ``out of plane" to produce features such as huts, pyramids and domes in GeSi/Si~\cite{Mo1990,Medeiros1998,Wiebach2000,Aqua2013} and InGaAs/GaAs films~\cite{Liu2000}. This points to the possibility that the direction of causation may go from phase separation towards  surface modulations, as opposed to the view implicit in theoretical studies of morphological instabilities in alloy thin films.

{\bf Acknowledgements}: AL and TAA thank the Department of Science and Technology of the Government of India for financial assistance through the TUE program. 

%

\end{document}